\documentclass[preprint,fleqn,showpacs,showkeys]{revtex4}
\usepackage{graphicx}
\usepackage{amssymb}
\usepackage{amsmath}
\usepackage{bm}
\begin{document}
\setcounter{page}{0}
\title[]{Long Journey toward the Detection of Gravitational Waves and New Era of Gravitational Wave Astrophysics}
\author{Hyung Mok \surname{Lee}}
\email{hmlee@kasi.re.kr}
\affiliation{Korea Astronomy and Space Science Institute, 776 Daedeokdae-ro, Daejon, 34055}

\date[]{Received 5 September 2018}

\begin{abstract}
The gravitational waves were detected directly for the first
time on September 14, 2015 by two LIGO detectors at Livingston, Louisiana and Hanford, Washington, USA. Careful analysis revealed that this signal was produced by the last moment of inspiral and
merger of two black holes that have been orbiting each other. Since the first detection, several other gravitational wave sources, including one neutron star merger event in August 201,7 have been observed by the LIGO together with Virgo. This article provides a very brief
overview of the history toward the understanding the nature of gravitational waves, detectors, types of sources and  achievements of the gravitational wave detectors.  We then discuss astrophysical implications of the recent detections of black hole and neutron star binaries  and provide short term and long term prospects of gravitational wave astronomy.
\end{abstract}



\maketitle

\section{Introduction}
Gravitational waves (GWs) were directly detected for the first time on September 14 by two LIGO  (Laser Interferometer Gravitational-wave Observatory) detectors located in Livingston, Louisiana and Hanford, Washington\cite{GW150914}. More events have been detected subsequently during two observing runs of the advanced LIGO, first one\cite{PRX-O1} from September 2015 to January 2016 and the second one from November 2016 to August 2017. Advanced Virgo joined the observing run in August 2017. The detected sources so far include mergers of six black hole binaries and one neutron star binary. The GW sources are designated by the dates of the detections: GW150914\cite{GW150914}, GW151226\cite{GW151226}, GW170104\cite{GW170104}, GW170608\cite{GW170608}, GW170814\cite{GW170814},  GW170817\cite{GW170817}. In addition there was a source with relatively low signal to noise ratio designated as LVT151012. All the signals except for GW170817 are interpreted as the gravitational waves  due to the merging of compact binaries composed of black holes. On the other hand GW170817 is due to the merging of a binary composed of two neutron stars. Electromagnetic radiation at various wavelengths has been observed after GW170817 by many observatories, including the ones operated by Seoul National University and Korea Astronomy and Space Science Institute (KASI).

Gravitational waves were predicted by Einstein in 1916\cite{gw_predict} by linearizing the field equation that he himself has proposed in 1915. It took more than 100 years to detect the natural phenomena after the prediction. This can be compared with the case of electromagnetic radiation which was theoretically predicted by James Clerk Maxwell in  1873\cite{maxwell} and detected experimentally by Heinrich Hertz in 1887. As we know, gravity is a very weak force and therefore gravitational waves were much more difficult to detect than the electromagnetic waves. Furthermore, the conceptual understanding of the gravitational waves was very difficult. We are familiar with the Newtonian law of the gravity, but gravitational waves are not generated in Newtonian theory of gravitation. General relativity (GR)  is a much more complex theory and often the conceptual understanding of the GR requires more imagination than Newtonian theory. Consequently, the confusion regarding the physical nature of the gravitational waves ensued for several decades after the prediction. The physical reality of the gravitational waves was widely accepted only in 1950s.

In the early days of the GW experiments, most notably led by Joseph Weber \cite{Weber_69} it was assumed the supernova explosion as the main source of GW. However, the GW are emitted when there is a significant acceleration of the quadrupole moment while stars remain nearly spherically symmetric until the last moment of explosion. Therefore the amplitude of the GWs expected from supernova explosion are not strong (see \S 5-B). Even with a substantial uncertainty, the detection is assumed to be possible only for those in the Galaxy. Moreover the supernova explosion rate per galaxy is expected to be rather low (current estimation is 1 per 50 yr \cite{al_abundance}). 

The existence and the effect of the GWs were confirmed by the monitoring of the orbital parameters of the binary millisecond pulsar PSR B1913+16 \cite{pulsar_confirm} that was discovered by Hulse and Taylor \cite{hulse_taylor}. The orbital period was found to change according to the prediction of energy and angular momentum loss by the gravitational waves. It should be noted that the discovery of the binary pulsar provided a very good motivation for the construction of the gravitational wave detector. 

PSR B1913+16 emits gravitational waves with very low amplitude at very low frequency ($\sim$70 $\mu$Hz) which is too low to be detected by the ground-based detectors. However, it will coalesce in about 300 million years. In fact most of the gravitational waves are emitted during the last moment of the coalescence and therefore coalescence of binaries composed of compact stars such as black holes or neutron stars are ideal targets for the ground based detectors.  However a chance of a merging event occurring in our Galaxy is one per hundred million years. Since the radio telescopes do not have enough sensitivity to detect all millisecond pulsars in the Galaxy, the expected rate should be somewhat higher. Furthermore, not all pulsars can be detected even in small distances since the pulsar emissions are highly beamed and not all observers can see the pulses.  More binaries composed of neutron stars have been discovered \cite{Wolszczan_91}\cite{Burgay_etal_06}\cite{Lyne_etal_04}\cite{Faulkner_etal_05} and subsequent studies taking into account the effects of sensitivity and the beaming demonstrated that the merging event could take place in our Galaxy once every few hundred housands to a few million years \cite{chunglee}. Since the number density of the galaxies similar to the Milky Way Galaxy is about $5\times 10^{-3} ~ {\rm Mpc}^{-3}$\cite{mweg_den}, gravitational wave detector sensitive enough to detect to the binary neutron star merger at about hundred Mpc would be able to make frequent detections of such events. Indeed the LIGO has detected the neutron star merger event on August 17, 2017 during its second observation run \cite{GW170817}. 

The existence of the black hole binaries has been speculated based on the theory of the stellar evolution and population synthesis. However, direct probe for the black hole binary based on electro-magnetic observation is not possible since black holes do not emit any radiation other than GWs. Several binaries composed of a black hole and a normal star have been identified from the X-ray observations  and the masses of the black holes have been measured from the orbital motion of the companion.  Typical masses of the black holes in X-ray binaries lie between 5 to 15 ${\rm M}_\odot$ \cite{x-ray_bh_mass}, much greater than the typical mass of neutron stars of about 1.4 ${\rm M}_\odot$. Since the amplitude of the gravitational waves at a given distance is approximately proportional to the mass of the binaries, the black hole binary merger can be seen at much greater distances than neutron star binaries. However, the rate of mergers of black hole binaries was much more uncertain than that of neutron star binaries since no black hole binary system had been observed until the detection of gravitational waves by the LIGO in 2015. As mentioned earlier, LIGO and Virgo have detected more gravitational wave events from black hole binaries than neutron star binaries. This is a surprise, although there have been some predictions of predominance of black hole binaries among the expected sources from the advanced LIGO/Virgo \cite{bae_etal_14}.

The present review is intended to give a general overview of the gravitational wave research and future perspectives. We start  from  history  of our understanding of the gravitational waves (\S2)  and the development of detectors (\S 3).  We then briefly review the propagation and generation  of gravitional waves in \S 4. \S 5 is devoted to the types of astrophysical sources and the methods for the data analysis and the overview of the observed sources and their astrophysical implications are discussed in \S 6. We further discuss the concept of the proposed detector in mid-frequencies, including one from Korean community based on superconducting instrumentation in \S 7. The final section summarizes.

\section{The Birth of the Concept of the Gravitational Waves}

The gravitational wave is a direct consequence of Einstein's general theory of relativity which was first announced in 1915. Einstein subsequently published two papers on the gravitational waves in 1916\cite{ gw_predict} and 1918\cite{gw_predict2}: the first one concerned with the propagation of the gravitational waves in vacuum while the second one entitled ``On the Gravitational Waves'' described how gravitational waves are generated.

However, Einstein was quite pessimistic about the possibility of detection of the gravitational waves experimentally because of the extreme weakness of the signal. The strength of the gravitational waves are expressed in the relative change of the length of two fixed points. Apparently, the gravitational waves did not receive much attention among scientists for many years. 

In  the middle of 1930s, an interesting event happened regarding gravitational waves. Nathan Rosen, a young physicist became an assistant of Einstein at the Institute for Advanced Study in Princeton, New Jersey. He carried out many important works with Einstein during his stay in Princeton until 1936 including the discovery of worm hole solutions in the general relativity.  He also persuaded Einstein that the gravitational wave does not exist. They wrote a paper entitled ``Do Gravitational Waves Exist?'' and submitted it to the Physical Review. In a letter to Max Born, Einstein wrote `` Together with a young collaborator, I arrived  at the interesting result that gravitational waves do not exist,  though they had been assumed a certainty to the first approximation. This shows that the non-linear general relativistic field equations can tell us more or, rather, limit us more than we 
have believed up to now.'' 

The referee's report  was very critical. A few days after Einstein received the report, he wrote an angry reply to the journal:

Dear Sir,

We (Mr. Rosen and I) had sent you our manuscript for publication and had not authorized you to show it to specialists before it is printed. I see no reason to address the - in any case erroneous - comments of your anonymous expert. On the basis of this incident I prefer to publish the paper elsewhere.

Respectfully,

P.S. Mr. Rosen, who has left for the Soviet Union, has authorized me to represent him in this matter.''

Einstein then submitted this paper to the {\it Journal of the Franklin Institute} without changes and it was accepted by the journal in its original form. However, the conclusion of the published version was opposite to the original version, which is not available today, even with change of the title to ``On Gravitational Waves''. What has happened between the submission of the paper and eventual publication in the {\it Journal of the Franklin Institute} is described in detail in an article by Daniel Kennefick\cite{Kennefick_05}. We simply note that Howard Percy Robertson, a professor at Princeton University at that time, persuaded Einstein that he had a mistake in the original paper. Nathan Rosen did not have any role in this process and he was not happy about Einstein's sudden change.
 
Most of the confusion around the gravitational waves was due to the choice of the coordinate systems and the interpretation of the results in terms of measurements. In fact the mistake by Einstein and Rosen was also caused by the singularity of the coordinate system. In 1956, Felix Pirani, a young scientist from Ireland, has published a very important paper in a Polish journal {\it Acta physica Polonica} (translated version:  \cite{Pirani_09}). In this paper, Pirani presented a mathematical formalism for the deduction of observable quantities of the gravitational waves. Pirani's work was presented in a conference held in 1957 at Chapel Hill, North Carolina, now known as {\it Chapel Hill conference}. This meeting was organized to commemorate an inauguration of the Institute of Field Physics (IOFP), whose purpose would be pure research in the gravitational fields in January 1957. The theme of the conference was ``The role of gravitation in Physics''. The conference was attended by mostly young physicists including Felix Pirani.  

Pirani has presented his idea on the effects of the gravitational waves.  He showed that a set of  freely falling particles would experience relative motions with respect to one another. At the end of the conference, Richard Feynman remarked that ``I think it is easy to see that if gravitational waves can be created they can carry energy and can do work. Suppose we have a transverse-transverse wave generated by impinging on two masses close together. Let one mass   carry a stick which runs past touching the other. I think I can show that the second in accelerating up and down will rub the stick, and therefore by friction make heat.'' This became known as {\it sticky bead} argument and  was used to explain the reality and existence of the gravitational waves \cite{DeWitt_Rickles_11}. 

Inspired by Pirani's talk, Joseph Weber, together with John Archibald Wheeler wrote a paper within three weeks of the Chapel Hill conference on the experimental ideas of gravitational waves. He began constructing the well known bar detector at the University of Maryland. He developed the first bar detector in early 1960s. When the gravitational wave passes the bar detector, which has a cylindrical shape with diameter of 1 meter and length of 2 meters, it experiences the tidal force that distorts the detector (eq(15) and eq(16)). The aluminum bar would respond most effectively by a gravitational wave of resonant frequency of 1660 Hz. The measurement error in this type of detector was about $10^{-16}$ m, corresponding to the minimum strain amplitude, which is a fractional change of length due to GWs (see eq(14), of  $10^{-16}$. As we will see in \S V typical astrophysical sources would have much smaller amplitude than this (by about 6 orders of magnitude) so that the detection of the gravitational waves with this type of the detector was not possible. He placed two detectors at large distance: one in the University of Maryland and the other in Argonne National  Laboratory which is 1,000 km away from the University of Maryland. He rightfully attempted to find coincidence signals in order to reject any noise due to environmental effects. He eventually claimed to have detected gravitational wave signal in 1969 \cite{Weber_69}. Here is the abstract of the paper entitled ``Evidence for Discovery of Gravitational Radiation'':

Coincidences have been observed on gravitational-radiation detectors over a base line of about 1000 km at Argonne National Laboratory and at the University of Maryland. The probability that all of these coincidences were accidental is incredibly small. Experiments imply that electromagnetic and seismic effects can be ruled out with a high level of confidence. These data are consistent with the conclusion that the detectors are being excited by gravitational radiation.

In retrospect, the understanding of the gravitational waves in late 1960s was very poor. Weber presumed that either supernova or pulsar is a promising source of the gravitational waves. The amplitudes of the gravitational waves coming from the supernova explosion is still very uncertain even today, but we know that even a supernova explosion within the Milky would require a detector with sensitivity of 6 orders of magnitude higher than Weber's bar detector. 

\section {The Birth of the Laser Interferometer}

After the announcement of the measurement of gravitational waves by Weber in 1969, several groups in the world tried to confirm by constructing independent detectors. However, nobody was able to reproduce the detection and the skepticism grew. 

Rainer Weiss at MIT was interested in the detection of the gravitational waves, but he was not able to understand Weber's papers. In fact, extraction of the gravitational wave amplitude from the electric signal of the bar detector is a complicated matter. Weiss wanted to have a straightforward experiment set up. Apparently he read papers by Pirani who demonstrated that the distance between two freely falling bodies can be changed by the gravitational waves. The idea of using laser interferometer to measure the gravitational waves occurred to a few scientists, including Weber in late 1960s.
Weiss recalled that the idea was incubated in 1968 or 1969 when he was teaching general relativity in MIT in an interview done as a part of a series of 15 oral histories conducted by the Caltech Archives and the entire text is available from \cite{weiss_interview}.  It was the time when Weber was publishing series of papers on the development of the bar detectors and results of measurements including the claim of detection.  

Weiss said in the above mentioned interview that he followed the discipline of Robert Dicke who was a supervisor of Weiss when he was a post-doc in Princeton University during 1962-1963. According to Weiss, Dicke  made all kinds of considerations before actually building experiments. 
Weiss sat in a little room and thought of every conceivable noise source he can think of. He then put these results in the Research Laboratory of Electronics quarterly report in 1972\cite{weiss_rle}. His conclusion was that Michelson laser interferometer with kilometer scale would be able to measure the gravitational waves of strain $h=\Delta L/L \sim 10^{-21}$ which is about 5 orders of magnitude better than Weber bar. The schematic drawing of the detector is shown in Fig. 1: this is the basis of the LIGO detector today. Weiss went on to build a 1.5 m scale prototype of laser interferometer and demonstrated the principle. 

Rainer Weiss also played an important role in measuring the cosmic microwave background and its spatial fluctuations. Since this article concerns mostly gravitational waves, we do not dwell too much on the cosmic microwave background, but the RLE reports n 1972 \cite{weiss_rle} was composed of two parts: Part A Balloon measurements of far infrared background radiation and Part B Electromagnetically coupled broadband gravitational antenna. 

\begin{figure}
\includegraphics[width=10.0cm]{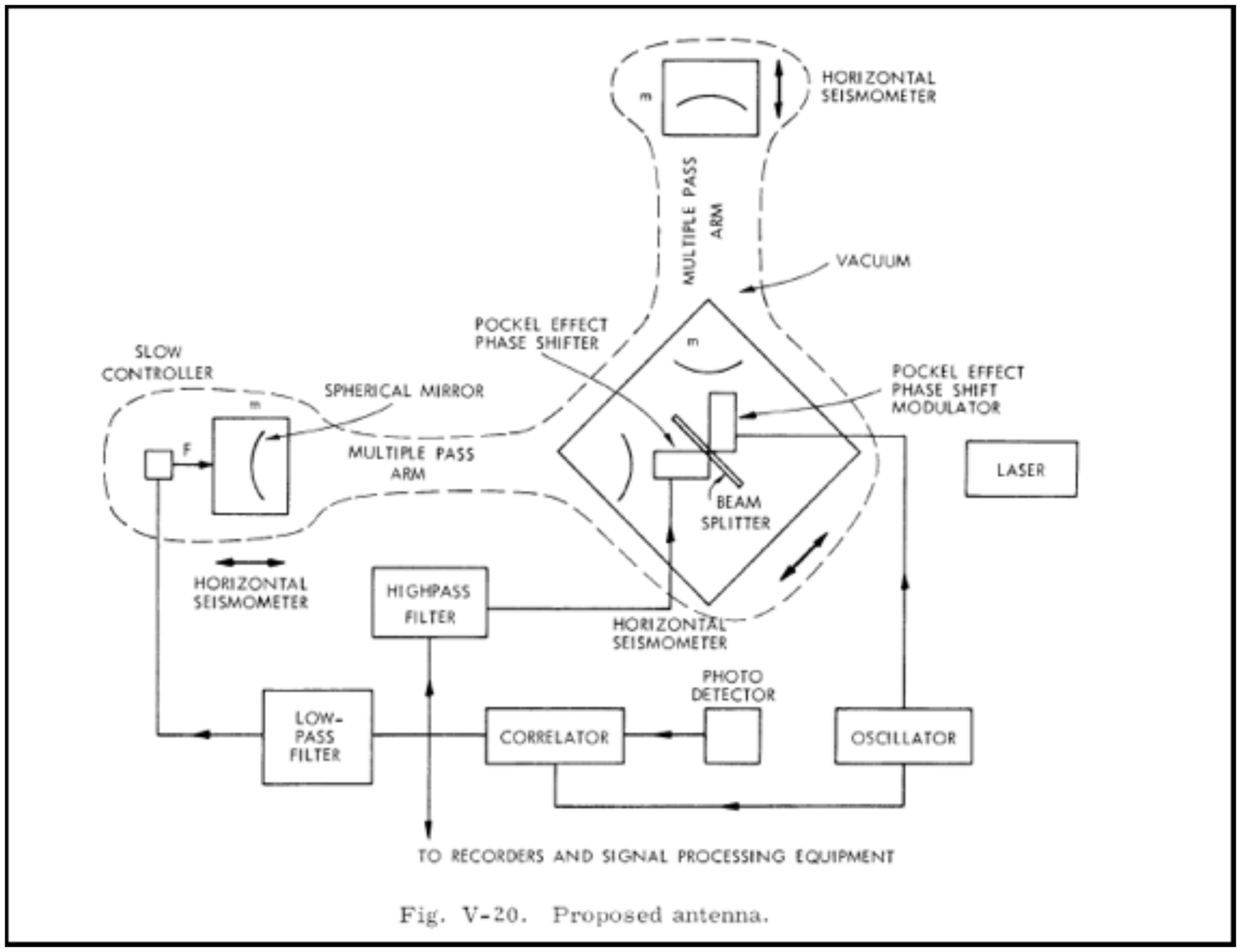}
\caption{The schematic drawing of laser interferometer gravitational wave antenna by Weiss.
The figure was taken from { https://dspace.mit.edu/handle/1721.1/56271}. }
\end{figure}

In the mean time, Kip Thorne became a member of faculty at Caltech in 1967. As a theoretical physicist and astrophysicist, Thorne was very interested in the astronomical aspects of the gravitational waves. Together with William Press, Thorne wrote a review article entitled ``Gravitational Wave Astronomy'' in Annual Review of Astronomy and Astrophysics \cite{Thorne_Press_72}. At that time, Thorne knew about the ideas of laser interferometer as a gravitational wave detector from Russian scientists, but he thought that the sensitivity of this type of detector is not sufficient. In fact, there is a following statement in the book `Gravitation' by Misner, Thorne, and Wheeler \cite{Misner_Thorne_Wheeler_73}
``interferometric detection isn't a practical way of doing gravitational wave astronomy''. 

Weiss heard about the review article by  Press \& Thorne before it was published and sent his calculation regarding the sensitivity of the laser interferometer to William Press who told about this to Thorne. Eventually, Press and Thorne put in the laser interferometer as a footnote of a new technique in the review.  

Thorne established a very strong theoretical group at Caltech since he became a faculty member and was thinking about `What should Caltech do as in an experimental way in gravitation'. Rainer Weiss did not know well about Thorne until they met in Washington DC. Weiss was charged to lead a committee on applications of space for gravitation by NASA and he invited Thorne to be a witness for the committee. They met in downtown Washington DC without hotel reservations. They shared a hotel room and had a discussion almost whole night. Apparently Weiss was able to persuade Thorne that the laser interferometer is the ultimate method to detect gravitational waves. 

Weiss received a small grant from NSF. Ronald Drever at Glasgow was recruited by Caltech to lead the laser interferometer project, initially as a part time member in 1979 and later full time in 1984. 

In 1989, a proposal for LIGO construction was jointly submitted by Caltech and MIT to the NSF which approved the project in 1990. The funding started in 1991 after the appropriation by the Congress in 1991. The Hanford, Washington and Livingston, Louisiana were chosen as the sites for the LIGO and the construction began in 1994 and the inauguration of the LIGO took place in 1999. Design sensitivity of the initial LIGO was reached in 2006. Initial LIGO operations have been done until 2010 and the advanced LIGO installation began in that year. The advanced LIGO installation was completed in 2014 and the first observing run of the advanced LIGO started in September 2015.


\section{Generation and Propagation of Gravitational Waves}

According to the Einstein's theory of general relativity, the geometrical structure of the space time is related to the 
energy and momentum of matter and radiation, according to the following field equation,

\begin{equation}
R_{\mu\nu}-{1\over 2} R g_{\mu\nu} =  {8\pi G\over c^4} T_{\mu\nu},
\end{equation}
where $R_{\mu\nu}$, $R$, $g_{\mu\nu} $ and $T_{\mu\nu}$ are Ricci tensor, curvature scalar, metric tensor and 
energy-momentum tensor, respectively,  $c$ is the speed of light, and $G$ is the gravitational constant. The Ricci tensor is related to the space-time curvature tensor and the curvature scalar is a real number determined by the intrinsic
geometry of the space-time manifold. 

If the gravitational field is weak, the metric tensor can be decomposed into two parts: the flat part and the small perturbation, i.e.,

\begin{equation}
g_{\mu\nu} = \eta_{\mu\nu} + h_{\mu\nu},
\end{equation}
where $\eta_{\mu\nu}$ is the Minkowski metric and $|h_{\mu\nu}| \ll 1$. 

Under such a  weak field limit, the Einstein's field equation becomes

\begin{equation}
 \square  h_{\mu\nu} - {\partial^2\over \partial x^\lambda\partial x^\mu} h^{\lambda}_\nu - {\partial^2\over \partial x^\lambda\partial x^\nu} h^{\lambda}_\mu + {\partial^2\over \partial x^\mu\partial x^\nu} h^\lambda_\lambda = -{16\pi G\over c^4} S_{\mu\nu},
\end{equation}
where 
\begin{equation}
S_{\mu\nu} \equiv T_{\mu\nu} - {1\over 2} \eta_{\mu\nu} T^\lambda_\lambda
\end{equation}
and
\begin{equation}
\square = -{\partial^2\over c^2 \partial t^2} + \nabla^2
\end{equation}
After performing coordinate transformation
\begin{equation}
x^\mu \rightarrow x^{\prime\mu} = x^\mu +\xi^\mu
\end{equation}
Then,
\begin{equation}
h^\prime_{\mu\nu} = h_{\mu\nu} - {\partial \xi_\mu\over \partial x^\nu} - {\partial \xi_\nu\over \partial x^\mu}.
\end{equation}
Now we further define a trace-reversed perturbation by
\begin{equation}
{\bar h_{\mu\nu}} =  h_{\mu\nu} - {1\over 2} h \eta_{\mu\nu}
\end{equation} 
and impose the Lorentz (or transverse) gauge condition of ${\partial{\bar h}_{\mu\nu}\over \partial x_\nu}=0$ 

\begin{equation}
\square {\bar h_{\mu\nu}} = -{16 \pi G\over c^4} T_{\mu\nu} .
\end{equation}
The above equation is a usual wave equation with a source term.  In vacuum (i.e., $T_{\mu\nu}=0$), it can be simply written $ \square {\bar h_{\mu\nu}} = 0$, and restricted gauge transformations further allow $\bar{h}^{\mu}_{\mu}=0$ (Traceless) and $\bar{h}_{0i}=0$. One can also have $\bar{h}_{00}=0$ by using the field equation for it in the absence of source and good behavior at infinity.  A plane wave solution can be written in the form

\begin{equation} 
{\bar h_{\mu\nu}} = Re \left\{ A_{\mu\nu} \exp(i k_\lambda x^\lambda)\right\} .
\end{equation}
The metric perturbation becomes very simple in transverse-traceless (TT) gauge. Furthermore, in TT coordinates ${\bar h_{\mu\nu}}=h_{\mu\nu}$.  Since we have a freedom to choose 
the direction of the propagation, the gravitational wave traveling along the $z$-direction can be written,

\begin{equation}
h^{TT}_{\mu\nu} = h_+ (t-z)  \begin{pmatrix} 
0 & 0 & 0 & 0\\
0 & 1 & 0 & 0\\
0 & 0 & -1 & 0 \\
0 & 0 & 0 & 0
\end{pmatrix}
+ h_\times  (t-z) \begin{pmatrix} 
0 & 0 & 0 & 0\\
0 & 0 & 1 & 0\\
0 & 1 & 0 & 0 \\
0 & 0 & 0 & 0
\end{pmatrix}
.
\end{equation}
In these coordinates, the line element becomes
\begin{equation}
ds^2 = -dt^2 + (1+h_+ ) dx^2 + (1-h_\times) dy^2 + dz^2 + 2 h_\times dx dy .
\end{equation} 
The lengths in x- and y-directions for $h_+$ between the origin ($x_1, y_1$) and the x-end ($x_2, y_1$) and y-end ($x_1, y_2$) points  then change in the following manner,
\begin{equation}
L_x = \int_{x_1}^{x_2} \sqrt{ 1+ h_+} dx \approx \left( 1+ {1\over 2} h_+  \right) L_{x0} ; \\
L_y = \int_{y_1}^{y_2} \sqrt{ 1- h_+} dy \approx \left( 1-{1\over 2}h_+   \right)L_{y0}  ;
\end{equation}
where $L_{x0}$ and $L_{y0}$ are the lengths of x- and y-arms in the absence of the gravitational waves (i.e.,$ h_+ =0$). The laser interferometers detect the gravitational waves by measuring the relative changes of the lengths of the two arms.  

Another possibility of measuring gravitational waves is to utilize the dynamical effects. In the presence of the gravitational forces, the two points separated by ${\bf\delta x}$ would experience the tidal force,

\begin{equation}
{d^2 \delta x_j\over dt^2} = -R_{j0k0} x^k = {1\over 2} {\ddot h}_{jk} x^k,
\end{equation}
where the Riemann tensor component $R_{j0k0}$ is a gravity gradient tensor in Newtonian limit, i.e.,
\begin{equation}
R_{j0k0} = -{\partial^2 \Phi\over \partial x^j \partial x^k} .
\end{equation} 
The diagonal components  of the gravity gradient tensor are the tidal forces and these forces can cause relative motions of two freely falling bodies or resonant motions of the metallic bar. In fact the first generation gravitational wave detectors by Joseph Weber tried to detect the piezoelectric effects of a metallic bar caused by the tidal forces acting on the bar (e.g.,\cite{Weber_60} ).  A tensor detector using superconducting devices recently proposed \cite{Paik_etal_16}  also tries to directly measure the tidal forces acting on the test masses which are magnetically levitated above a rigid body platform. 

\section {Gravitational Wave Sources}

The gravitational wave generation can be understood by looking at the equation (18) more closely. In the presence of the sources of the gravitational force, the wave equation should include the source term, i.e.,
\begin{equation}
\square {\bar h}_{\mu\nu} = -{16 \pi G\over c^4} T_{\mu\nu}.
\end{equation}
which is similar to the Poisson's equation in the Newtonian limit. The formal solution can be written as
\begin{equation}
{\bar h}_{\mu\nu} = {4G \over c^4} \int{T_{\mu\nu} (t-|{\bf x} - {\bf x}^\prime|, {\bf x}^\prime) \over |{\bf x}-{\bf x}^\prime|} d^3 x^\prime \approx {4G\over rc^4} \int T_{\mu\nu} (t-|{\bf x} - {\bf x}^\prime|, {\bf x}^\prime) d^3x^\prime,
\end{equation}
where we used the fact that the observer is located at a distance $r$ which is much greater than the size of the source (i.e., range of integration of the variable $x^\prime$). Notice also that the integral is evaluated at the retarded time, i.e., $ct  \rightarrow ct-|x-x'|$.The energy-momentum conservation equation $T_{\mu\nu;\mu}=0$, where $;$ represents a covariant derivative, can be written as
\begin{equation} 
{\partial T^{tt}\over \partial t} + {\partial T^{kt} \over \partial x^k}=0,
\end{equation}
which can be differentiated once more with respect to time to obtain
\begin{equation}
{\partial^2 T^{tt} \over \partial t^2} = {\partial^2 T^{kl}\over \partial x^k \partial x^l}.
\end{equation}
Here $\partial_t T^{tk} +\partial_l T^{lk}=0$ for $\nu=k$ is used. One can show that the strain amplitude can be expressed as follows by multiplying $x^j x^k$ to the equation above with integrations over the spatial volume,
\begin{equation}
{\bar h}_{jk} = {2G\over rc^4} {d^2 Q_{jk} (t-r)\over dt^2} 
\end{equation}
where $Q_{jk} = \int \rho x^j x^k d^3 x$ is the moment of inertia tensor of the source or is often referred to as the quadrupole moment tensor. 

The second-order time derivative of the quadrupole moment is regarded as the 'quadrupole kinetic energy', i.e.,
\begin{equation}
{\ddot Q}_{jk} \sim {{\rm(mass)}\times {\rm (size)}^2 \over ({\rm transit ~ time})^2}= \epsilon Mc^2
\end{equation}
In most cases $\epsilon$ is much smaller than 1 and therefore the gravitational waves are weak, but it could become almost 0.1 in special cases. For astronomical sources with mass $M$ at distance $r$ amplitude $h$ can be estimated 
\begin{equation}
h_{jk} \sim 10^{-22} \left({\epsilon \over 0.1}\right) \left({M\over M_\odot}\right) \left({100 {\rm Mpc}\over r}\right).
\end{equation}

The above equation tells us the reason why it is so difficult to detect gravitational waves. First of all, in order for $\epsilon$ to be of order of 0.1, the objects radiating gravitational wave should be very compact. Otherwise, the maximum value for the $\epsilon$ is only of order of $v_{esc}/c^2$. For the case of the ordinary stars, $v_{esc}$ is  a few hundred km/s, resulting $\epsilon_{max}\sim 10^{-9}$. The best candidates of the gravitational wave sources are black holes or neutron stars.  Furthermore, the object should have sufficiently large quadrupole moment.  Single neutron stars or black holes are likely to produce negligible amount of gravitational waves   compared to binary stars.

The gravitational sources are divided into the following four categories: compact binary coalescence, continuous, burst and stochastic sources. These sources are described below.

\subsection{Compact Binary Coalescence}
As we can see from the eq. (21), the sources with large amount of time varying quadrupole moment would produce the strongest gravitational waves. The binary stars composed of two compact stars such as black holes and/or neutron stars satisfy such condition. The orbital motion of a binary system naturally produces gravitational waves which carry away both energy and angular momentum. The orbital evolution of binary system has been studied in detail by \cite{Peters_64}  in the context of post-Newtonian (PN) approximation. As the orbital energy and angular momentum becomes smaller, the orbital radius and eccentricity become smaller. Eventually the binary stars coalesce and form a single star. For the case of black holes, it is likely that there will not be any ejection of material and therefore the outcome of a merger of black hole binary is another black hole with greater mass. However, the neutron star merger involves hydrodynamical interactions and therefore would accompany with the electromagnetic radiation.

The early evolution of the binary evolution can be approximated by the series of Keplerian motion which can be specified by the eccentricity ($e$) and the semi-major axis ($a$). The amount of energy carried by gravitational waves emitted from the binary motion depends sensitively on the orbital radius. Therefore, the binary coalescence is likely to be an accelerating process, with very slow change of the orbital radius until the last moment of the merger. For binaries composed of non-spinning components the post-Newtonian approximation up to order $\left(v/c\right)^2$ in orbital motion and up to order $(v/c)^5$ in energy and angular momentum loss give the following expression for orbit-averaged rates of changes of $a$ and $e$,

\begin{equation}
\left<{da\over dt}\right> = -{64\over 5} {G^3 m_1 m_2 (m_1 + m_2)\over c^5 a^3 (1-e^2)^{7/2}} \left( 
1+ {73\over 24} e^2 +{37\over 96} e^4\right)
\end{equation}
\begin{equation} 
\left<{de\over dt}\right> = -{304\over 15} e {G^3 m_1 m_2 (m_1 + m_2)\over c^5 a^4 (1-e^2)^{5/2}} \left( 
1+ {121\over 304} e^2 \right) ,
\end{equation}
where $m_1$ and $m_2$ are the masses of the binary components. In the presence of spin, the orbital evolution is much more complicated toward the final phase of the coalescence. The above formula is valid for sufficiently small $(v/c)$.

The time until final merger can be obtained by integrating the above equations from a given time when $(a,e)=(a_0,e_0)$ until $a\rightarrow 0$;
\begin{equation}
t_m={5\over 512} {c^5 a_0^4 (1-e_0^2)^{7/2}\over G^3 m_1 m_2 (m_1+m_2)}\left( 1+{73\over 24}e_0^2 + {37\over 96} e_0^4\right)^{-1}.
\end{equation}
As we can see from the above equation, the time until merge is very sensitive to $a_0$ and $e_0$, although to a much less extent. 

The orbital frequency of a Keplerian system with $(a,e)$ is $\omega=\sqrt{ G\mu/a^3}a^3$ where $\mu$ is the reduced mass of the binary system. Since the gravitational wave is the quadrupole radiation, the wave frequency is twice of the orbital frequency, i.e. $\omega_{GW} = 2\omega$. As the orbit shrinks, the GW frequency increases. The amplitude of the GW also increases as the oscillation frequency increases since the GW amplitude is proportional to the second derivatives of the quadrupole momenta. Such a signal that increases the frequency and amplitude sharply toward the final merger is called `chirp'. The phase with slow decrease of orbital separation and eccentricity is called inspiral, which lasts until the circular orbit becomes unstable. For the case of a test particle around a point mass, the circular orbit exists until $a = 3 r_s = 6 GM/c^2$ where $r_s$ is the Schwarzschild radius of a point mass $M$. In the case of arbitrary mass ratio binaries, the Innermost Stable Circular Orbit (ISCO) is slightly different from the above criterion and more careful calculations are necessary to locate ISCO for general two-body motions \cite{ISCO}. In any case the post-Newtonian approximation is not valid near the ISCO. Fully relativistic numerical simulations are necessary to compute the waveforms.

When two black holes become closer than the innermost stable circular orbit, they quickly merge into a single black hole. Just after the formation of a single black hole, the gravitational field settles into a static one after short phase of `ringdown' during which kind of damped harmonic oscillator-like GWs are emitted. Therefore, the complete waveform from compact binary coalescence can be described by the inspiral-merger-ringdown (IMR)  phases. An example of IMR waveform is shown in Fig. 2.

\begin{figure}
\includegraphics[width=10.0cm]{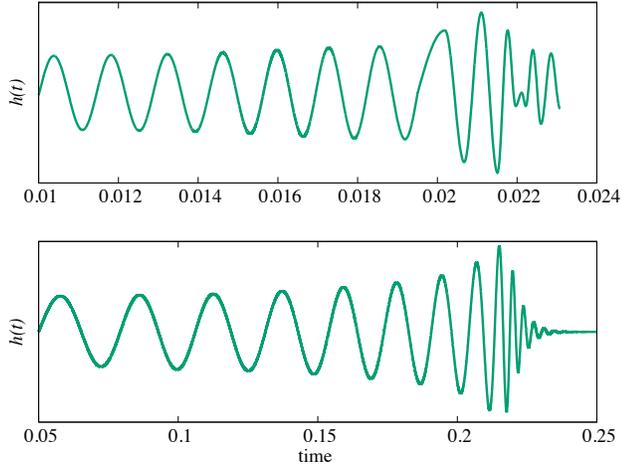}
\caption{Example of waveforms from neutrons star (upper) and black hole  binaries (lower). The numerical simulation data are  taken from \cite{Duez_etal_01}. 
}
\end{figure}

The waveform of the emitted gravitational waves by a merging black hole binary depends on the masses and spins of black holes for binaries in quasi-circular orbits. The observed waves further depends on the distance to the source, inclination of the binary, and the direction in the sky. In principle, these parameters can be determined by analyzing the observed waveforms. However, the inclination angle and the distance are degenerate: both quantities affect only the strain amplitudes. Furthermore, up to the 2.5PN order, the waveform depends on the mass only through the chirp mass which is defined as

\begin{equation}
{\cal M}={(m_1m_2)^{3/5}\over (m_1+m_2)^{1/5}} = \mu^{3/5} M^{2/5}
\end{equation}
where $M=m_1+m_2$ is the total mass and $\mu={m_1m_2\over m_1+m_2}$ is the reduced mass of the binary system. Of course such a degeneracy will be broken in the higher PN order, but the effect is generally small.  Such degeneracies mean that the there are large uncertainties in the estimated individual masses and the distances to the sources even for relatively strong gravitational wave sources.

The gravitational waves due to the coalescence of compact binaries can be best analyzed by using the matched filtering technique. Although the gravitational waves are described by a tensor quantity, the output from any detector is a time series of a scalar quantity. Such a relationship can be expressed by
\begin{equation}
h(t) = D^{ij}h_{ij}(t)
\end{equation}
where $D_{ij}$ is called detector tensor. However, the detector also produces its own noise which is denoted by $n(t)$. Therefore the actual readout from the detector is a combination of noise and the signal, i.e.,
\begin{equation}
s(t) = h(t)+n(t).
\end{equation}

If the signal $h(t)$ is significantly larger than noise $n(t)$, the detection of the signal is easy and straightforward. However, in general $|h(t) \ll |n(t)|$, we need to carry out a careful analysis. If we have knowledge of the shape of the signal itself, we can apply filtering technique to dig out the signal in the forest of noise. To illustrate this, let's multiply $s(t)$ by $h(t)$ and carry out the integration over some time $T$, which is the duration of the observation,
\begin{equation}
{1\over T} \int_0^T dt s(t) h(t) = {1\over T} \int_0^T dt h^2 (t) + {1\over T} \int_0^T dt n(t) h(t).
\end{equation}
We can see that the first integration of the above equation will grow with time, while the second integration does not as long as the noise and the signal are not correlated. If we observe sufficiently long time, the second term should go to zero and we can say the noise is filtered out. 

However, we do not have {\it a priori} knowledge on $h(t)$, but for the case of compact binary mergers we know that one of the  large number of waveforms with different parameter set should have an identical form of $h(t)$.  We need a procedure to find out the optimal filter that maximizes fitting the observed data best. In order for that we need a quantity that measures the goodness of the fit.  For that let's define
\begin{equation}
{\hat s} = \int_{-\infty}^\infty dt s(t) K(t)
\end{equation}
where $K(t)$ is called filter function.  The expected values of ${\hat s}$  for the case with signal and noise is denoted as $S$ while rms value of ${\hat s}$ in the absence of signal  (i.e., $s(t) =n(t))$ is the noise $N$). Since $<n(t)>=0$, $S$ becomes

\begin{equation}
S=\int_{-\infty}^\infty dt <s(t)> K(t) = \int_{-\infty}^\infty dt h(t) K(t) =\int_{-\infty}^\infty df {\tilde h}(f) {\tilde K}^* (f),
\end{equation}
and
\begin{eqnarray*}
N^2 &=& [<{\hat s}^2(t)> - <{\hat s}(t)>^2]_{h=0}=<{\hat s}^2(t)>  \\
&=&\int_{-\infty}^\infty dt dt^\prime K(t) K(t^\prime) <n(t) n(t^\prime)> \\
&= &\int_{-\infty}^\infty dt dt^\prime K(t) K(t^\prime) \int_{-\infty}^\infty df df^\prime e^{2\pi i ft - 2\pi i f^\prime t^\prime }<{\tilde n}^*(f) {\tilde n} (f^\prime)>.
\end{eqnarray*}
If the noise is assumed to be stationary $ <{\tilde n}^*(f) {\tilde n} (f^\prime)> =\delta(f-f^\prime) {1\over 2} S_n (f)$, where $S_n$ is the single sided noise power spectrum,
\begin{equation} 
N^2=\int_{-\infty}^\infty df {1\over 2} S_n(f) \{\tilde K(f)|^2.
\end{equation}
Therefore the signal-to-noise ratio $S/N$ becomes
\begin{equation}
{S\over N} = {\int_{-\infty}^\infty df {\tilde h}(f) {\tilde K}^* (f)\over [\int_{-\infty}^\infty df {(1/ 2)} S_n(f) \{\tilde K(f)|^2]^{1/2}}.
\end{equation}

Now we need to find out $K(t)$ that maximizes the S/N ratio. For convenience, let's define the inner product of two real scalar functions $a(t)$ and $b(t)$ with their Fourier transforms
 ${\tilde a}(f)$ and ${\tilde b}(f)$, respectively, as follows:
\begin{eqnarray*}
(a | b) &\equiv&{\rm Re}\int_{-\infty}^\infty df {{\tilde a}^*(f) {\tilde b}(f)\over (1/2) S_n(f) }\\
&=& 2{\rm Re}{\int_0^\infty df {{\tilde a}^*(f) {\tilde b}(f)\over (1/2) S_n(f)}}.
\end{eqnarray*}
The second part is due to the fact that  $a$ and $b$ are real functions so that ${\tilde a}(-f) = {\tilde a}^* (f)$.  Then the $S/N$ ratio simply can be written as
\begin{equation} 
{S\over N} = {(u|h)\over (u|u)^{1/2}}
\end{equation}
where $u(t)$ is the function whose Fourier transform is
\begin{equation}
{\tilde u}(f) = {1\over 2} S_n(f) {\tilde K}(f).
\end{equation}
The interpretation of the above result is very straightforward. The $S/N$ ratio can be maximized by making  ${\hat n}\equiv u/(u|u)^{1/2}$ parallel to $h$, like in case of vector. This can be done by choosing ${\tilde u} (f)$ proportional to ${\tilde h}(f)$, i.e.,
\begin{equation}
{\tilde K}(f) \propto {{\hat h}(f)\over S_n(f)}
\end{equation} 
so that 
\begin{equation}
{S\over N} = (h|h)^{1/2} = \left[ 4 {\int_0^\infty df {|{\tilde h}(f)|^2\over S_n(f)}} \right]^{1/2}.
 \end{equation}
Such a choice of the filter function of $K(t)$ is called matched filtering. For a given set of physical parameters for the compact binary coalescence, $h(t)$ can be computed. 
 
The above descrition gives how the signal-to-noise is obtained for an optimal filter, which requires certain model parameters. However, the presence of the noise means that there are some uncertainties in the estimated parameter set (see Refs \cite{det_char1} \cite{det_char2} \cite{det_char3} for detailed description of how detector noises are characterized). In that sense the parameter set that maximizes the $S/N$ may be the one with highest probability, but not the only one. Each parameter in the set could have some probabiity distribution. The probability distribution of each parameter can be obtained by the Bayesian statistics. 

According to Bayes theorem, the {\it posterior} probability of  parameters ${\vec \lambda}$ given the data $ s$ can be computed by
\begin{equation}
p({\vec \lambda}| s) \propto p( {\vec \lambda}| M) p(s| {\vec\lambda}, M)
\end{equation}
where $M$ is the model for signal and noise, $p({\vec x}| {\vec\lambda}, M)$ is the likelihood function that measures how well the data fits the model $M$ for the parameter set ${\vec\lambda}$, and $p{\vec\lambda})$ is the called $prior$ that measured the prior probability of model $M$.  Usually the priors are now well known and could be subjective. The likelihood function can be computed from

\begin{equation}
p(s|{\vec\lambda},M) \propto {\rm exp}\left(<s|h({\vec\lambda}> - {1\over 2} <h({\vec\lambda})|h({\vec\lambda})>\right).
\end{equation}
 For  compact binaries in circular orbit composed of point masses  (i.e., black holes) with spin, the number of parameters determining the wave form is 8: two for masses, 6 for spins since spin is a vector quantity. These are called intrinsic parameters. The observed signal is a projection of the gravitational wave onto the detector. For this projection, we need 7 more parameters: distance, time of the event, direction of the source (right ascension and declination), inclination angle of the orbital plane, reference time and the phase and polarization angle.  (see \cite{gen_text}). If we ignore spins, the number of parameters becomes 9. The numerical computation of the posterior probability distribution thus requires multi-dimensional integrals, for which Markov chain Monte Carlo (MCMC) method is found to be useful.  

\subsection{Continuous Sources}

The compact binaries produce gravitational waves at nearly constant frequencies when the separation is large since the dynamical effect is very small.  However, the frequency during such an early phase is usually much lower than the sensitive frequency range of the current ground-based GW detectors. On the other hand, some of the pulsars are known to have very short pulse periods (as short as $\approx$ milliseconds). These are rapidly rotating neutron stars with high degree of spherical symmetry. However, the neutron stars could have small irregularity on the surface and gravitational waves with nearly constant frequency could be emitted. If the neutron star is not axisymmetric, it will produce gravitational waves of frequency twice of the rotational frequency.  The rotation speed of neutron stars are known to be slowly decreasing. The maximum amount of gravitational waves expected from the rapidly rotating neutron stars can set the upper limits on the strain of the gravitational waves.

Consider a triaxial compact star, rotating around $z$-axis. The asymmetry of the star is parametrized by
\begin{equation}
\epsilon = {I_{xx}-I_{yy}\over I_{zz}} ,
\end{equation}
where $I_{xx}$, $I_{yy}$ and $I_{zz}$ are tmoments of inertia along $x-$, $y-$ and $z-$ axes, respectively. The expected strain amplitude of the gravitational wave when the rotation axis is pointing to the Earth  is

\begin{equation}
h_0= {4 \pi^2 GI_{zz} f_{GW}^2 \over c^4 r} \epsilon \approx 1.1\times 10^{-24} \left({I_{zz}\over I_0}\right) \left({f_{GW}^2\over 1  {\rm kHz}}\right)^2 \left( { {\rm kpc}\over r}\right)\left({\epsilon\over 10^{-6}}\right)
\end{equation}
where $I_0 = 10^{38} {\rm kg m}^2$ is the typical value of the moment of inertia of neutron stars. The total power emitted by the star in gravitational waves is
\begin{equation}
{dE\over dt} ={32\over 5} {G\over c^5} I_{zz}^2 \epsilon^2 \omega^6
\end{equation}
For the case of pulsars, the spin-down rates (${\dot f}$) are well measured. If we assume that the spin-down is entirely due to the gravitational radiation, the corresponding gravitational wave amplitude can be obtained by equating $dE/dt = I_0 \omega {\dot\omega}$\cite{riles_17},
\begin{align}
h_{sd} &= {1\over r} \sqrt{ -{5\over 2} {G\over c^3} I_{zz}{{\dot \omega}\over\omega}}\\
& 2.5\times 10^{-25} \left({ 1 {\rm kpc}\over r}\right)\sqrt{\left({1 {\rm kHz}\over f_{GW}}\right) \left({{\dot f}_{GW}\over 10^{-10} {\rm Hz/s}}\right)\left({I_{zz}\over I_0}\right)}.
\end{align}

Searches for gravitational waves from 200 pulsars using the data from the first observing run of the advanced LIGO have been carried out, but no significant  evidence for the gravitational wave signal from these pulsars has been found\cite{pulsar_search}. For eight  of these pulsars, their upper limits on $h_0$ are better than spin-down limits.

In facr neutron stars are likely to be losing their rotational energy through emission of electromagnetic pulses. 
This process pushes them to evolve more spherical shape while the crust is forced to stay their shape due to its hardness. Then the tension applied to the crust becomes
larger and larger as the neutron star rotates slower. At
some point, when the crust cannot retain its shape, it
breaks up with changing moment of inertia which leads
the changes in rotation velocity of the crust. However
such a model cannot explain the multiple pulsar glitches in
a decade due to its very slow energy losing mechanism.
They cannot accumulate enough energy that triggers the
glitch in a decade. The other model is vortex unpinning
of super fluid at the crust-core interface
 \cite{Packard_72}\cite{Anderson_Itoh_75}. Spin down rate of pulsar is much
smaller than the expectation rate when the neutron stars
consist of normal  fluid, therefore many theorists believe
that neutron stars' interior except the crust is made up
of super fluid. The difference in spin down rate between
core and crust causes differential rotation i.e. the crust
rotates slower than the core. This difference becomes
larger and larger and finally at some points when the
vortexes unpin, the core transfer the angular momentum
to the crust to make a uniform rotation. This causes a
glitch in rotation frequency. This model can explain well
the multiple glitches in some pulsars.

The gravitational wave expected from the pulsar glitch
is mainly due to the neutron star's normal modes excited by the glitch and shows the superposition of each
modes. Hence, it is necessary to have knowledge about
the non-radial pulsation modes of the rotating stars. If we can identify the normal modes that are
responsible for the gravitational waves from neutron stars just after the glitches, we will have very
strong constrains on the size of the neutrons stars since the frequencies of the normal modes 
are sensitive function of the density. Numerical models

\subsection{Bursts}
There are many explosive phenomena in the universe such as supernova and gamma-ray bursts. The supernova is induced by the runaway collapse of the stars to a black holes or neutron stars that inevitably rotate rapidly. If the collapse is spherically symmetric, no gravitational wave is expected. Observationally the deviation from the spherical symmetry is inferred from very high speed motion of neutron stars. However, such deviation from spherical symmetry during the collapse can arise due to various types of instability such as Standing Accretion Shock Instability (SASI). Strong magnetic field also could lead to the asymmetry of the collapse.  The gravitational wave amplitude is approximately

\begin{equation}
h_{ij}\sim {2G\over r c^4} {\ddot Q}_{ij} ,
\end{equation}
but we do not have accurate information on ${\ddot Q}_{ij}$. We simply  replace the second time derivatives of the quadrupole moment by the kinetic energy of the system with some factor $\alpha_Q$, i.e.,
\begin{equation}
h_{ij}\sim {2\alpha_Q G E_{kin} \over rc^4}.
\end{equation}

The generation of the gravitational waves mostly is thought to occur during the collapse and bounce phase. Gravitational waves can be emitted due to $g$-mode oscillation of the proto-neutron star. Numerical simulations are necessary in order to follow the detailed processes of the collapse, development of instability and resulting gravitational waves. The frequency of the gravitational waves expected from the supernova explosion lies between 100 to 1000 Hz while the strain amplitude at 10 kpc is estimated to be between $10^{-23}$ to $10 ^{-21}$ [see for example \cite{Ott_09}\cite{Morozova_etal_18}] .  Therefore, it is difficult to detect the  gravitational waves coming from the supernova in external galaxies. 

According to the grand unified theory the universe undergoes a series of phase transitions as the temperature drops during the early phase of the evolution. Cosmological defects such as cosmic strings are expected to be left behind after the phase transition.  The strings are expected to interact each other producing cusps. The burst of gravitational wave is expected in the vicinity of highly relativistic points on the strings.  The search for the burst signal from cosmic string on the LIGO data has been performed, but no direct detection has been made\cite{string_search}.

Hyperbolic encounters between two black holes also lead to the production of burst signa  during the periapse passagesl\cite{bae_etal_17}\cite{cho_etal_18}. The strength of the signals depends on the periapse distance with relatively simple shapes. The examples of the waveforms computed with 2.5 and 3.5 PN for non-spinning black hole pairs are shown in Fig. 3. Although the hyperbolic orbit generally produce only one burst per encounter, gravitational radiation capture can take place if the amount of the radiated gravitational waves is greater than the initial (positive) orbital energy as discussed in the next session. In such a case, the black holes return to the periapse repeatedly until the full circulatization. Therefore hyperboic encounters that lead to the gravitationacan be a potential burst sources for  the next generation gravitational wave detectors.

\begin{figure}
\includegraphics[width=10.0cm]{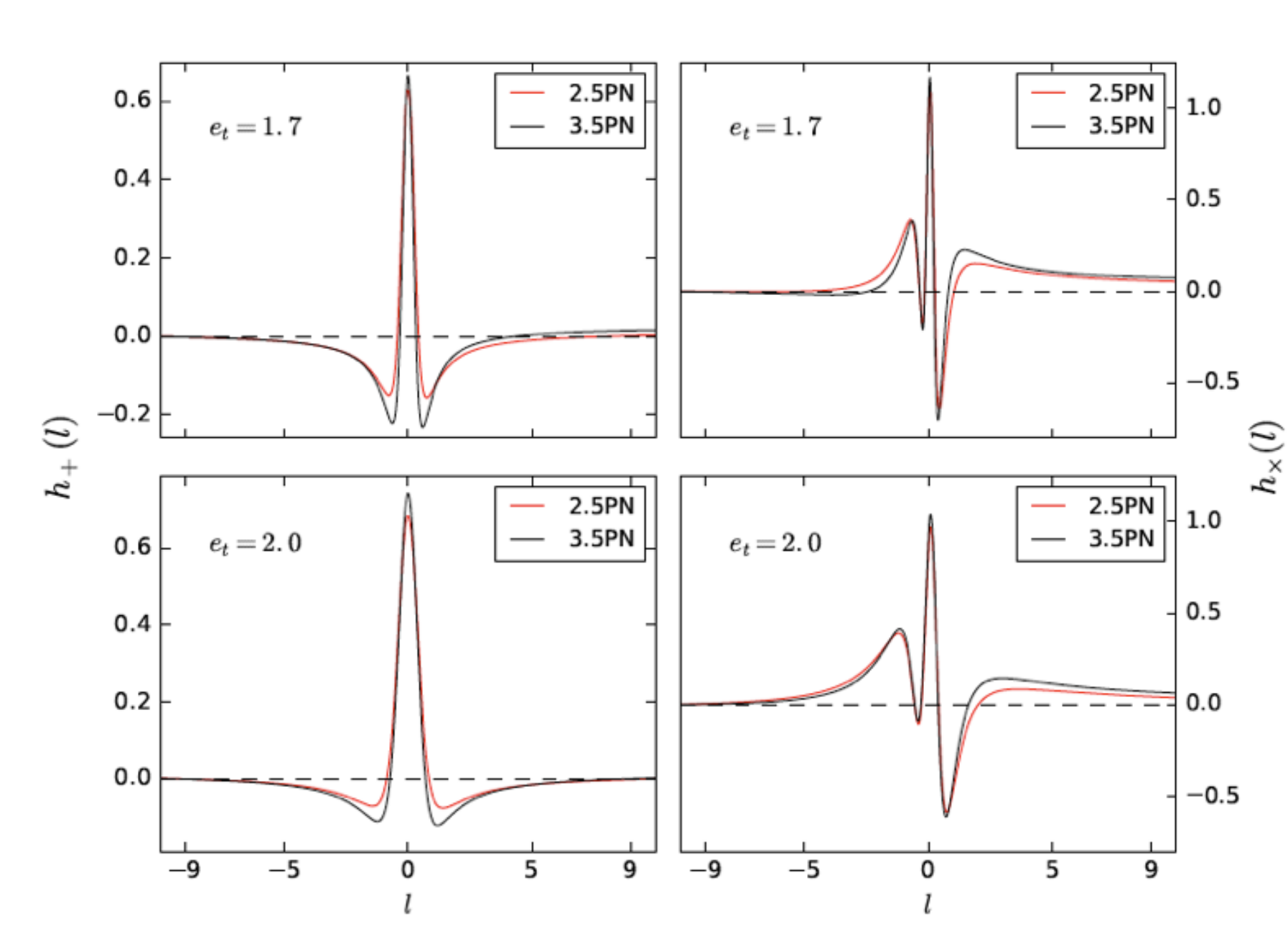}
\caption{Example of waveforms from during the periapse passages of two identical mass back holes with eccentricities shown in each panel. The red and black lines are using 2.5 and 3.5 PN approximation, respectively. Here $l$ is the normalized time centered on the time of peripase passage (Figure credit: Gihyuk Cho). 
}
\end{figure}

\subsection{Stochastic Background}

Stochastic background gravitational waves are the superposition of many incoherent sources. This is reminiscent of the background radiation in various wavelengths of electromagnetic waves. While the cosmic microwave background radiation is due to the black-body radiation at the time of recombination, background radiation at other wavelengths is mostly due to the unresolved sources such as distant galaxies. Similarly, there could be primordial gravitational waves emitted just after big bang since the universe is transparent for gravitational waves all the way to the big bang. However, mergers of neutron stars or black holes at very large distances cannot be resolved but form stochastic background gravitational waves. 

Since the stochastic background radiation is a superposition of waves with all possible propagation directions, the strain amplitude at time $t$ and position ${\bf x}$ can be formally written
\begin{equation}
h_{ij}(t ,{\bf x}) = \sum_{A=+,\times}\int d^2{\hat n}  \int_{-\infty}^\infty df {\tilde h}_A (f, {\hat n})e_{ij}^A ({\hat n})e^{-2\pi if(t-{\hat n}\cdot {\bf x}/c)}
\end{equation}
where ${\hat n}$ is the direction of propagation vector and $e_{ij}$ is the basis tensor of each polarization $A$ and ${\tilde h}_A (f, {\hat n})$ is the strain amplitude that characterizes the stochastic background. The stochastic background is assumed to be statio, Gaussian, isotropic and unpolarized.

For many copies of statistically identical signals, we can take ensemble average to obtain the physical quantities similar to statistical mechanics. However, we cannot take an ensemble average since there is only one universe. Instead we can take temporal average,
\begin{equation}
< h_{ij} h^{ij}> = \sum_{A, A^\prime =+,\times} \int_{-\infty}^\infty  \int_{-\infty}^\infty df df^\prime \int\int {d{\hat n}\over 4\pi}{d{\hat n}^\prime \over 4\pi}\left< {{\tilde h}_A}^* (f,{\hat n)}  {\tilde h}_{A^\prime}(f,{\hat n}^\prime )\right>.
\end{equation}
where spectral density $S_{GW}$ is defined through
\begin{equation}
\left< {{\tilde h}_A}^* (f,{\hat n)}  {\tilde h}_{A^\prime}(f,{\hat n}^\prime )\right> =\delta (f-f^\prime) {\delta^2({\hat n},{\hat n}^\prime) \over 4\pi} \delta_{A,A^\prime} {1\over 2} S_{GW}(f).
\end{equation}. 
Then the temporal average of the stochastic background becomes,
\begin{equation}
< h_{ij} h^{ij}> = 4 \int_0^\infty df S_{GW}(f)
\end{equation}
The stochastic background is characterized by the energy density. Since the energy-momentum tensor of gravitational waves is
\begin{equation}
t_{\mu\nu}= {c^4\over 32 \pi G} \left< \partial_\mu h_{\alpha\beta}\partial_\nu h^{\alpha\beta}\right>, 
\end{equation}
whose (0,0) component corresponds to the energy density, i.e.,
\begin{equation}
t_{00}={c^2\over 32\pi G} \left< {\dot h}_{ij}{\dot h}^{ij}\right>
\end{equation}
Therefore, the energy density carried by the stochastic background is 
\begin{equation}
\rho_{GW} ={c^2\over 32\pi G} \left< {\dot h}_{ij}{\dot h}^{ij}\right>
\end{equation}

Traditionally, the strength of the stochastic background is measured by the density parameter contained in gravitational waves so that
\begin{equation}
\Omega_{GW} = {\rho_{GW}\over \rho_{crit}}
\end{equation}
where $\rho_{crit}$ is the density required for a flat universe for a given Hubble parameter $H_0$, i.e.,
\begin{equation}
\rho_{crit} = {3 H_0^2 c^2\over 8\pi G}.
\end{equation} 
We further define the energy density in logarithmic frequency interval such that

\begin{equation}
\Omega_{GW} (f) = {1\over \rho_{crit}} {d\rho_{GW}(f) \over d\ln  f}.
\end{equation}
By inserting the plane wave ensemble expression into the energy density  in equation (48), we obtain
\begin{equation}
\rho_{GW} = {c^2\over 8\pi G} \int_0^\infty d\ln f f (2\pi f)^2 S_{GW}(f).
\end{equation}
By comparing eqs (51) and (52), we find the relationship between $\Omega_{GW} (f)$ and $S_{GW} (f)$ 
\begin{equation}
S_{GW} = {3 H_0^2\over 10\pi^2} f^{-3} \Omega_{GW} (f).
\end{equation} 

Now the isotropic stochastic background can be considered in a detector as an additional source of noise, and therefore identification of stochastic background is difficult unless we have a perfect knowledge of the detector noise. For the case of anisotropic sources, sidereal time modulation due to the relative motion of the detectors can be seen. However, we can apply technique similar to the matched filtering by computing the cross correlation between the outputs of two detectors as can be seen below.

Let's denote the output from the i-th detector as $x_i(t) = h_i (t) + n_i (t)$. Then the cross-correlation between two detectors over the observation period $T$ can be written
\begin{equation}
Y \equiv \int_0^T dt_1 \int_0^T dt_2 x_1(t_1) Q(t_1-t_2) x_2 (t_2),
\end{equation}
where $Q$ is a filter function which can be chosen to maximize the signal-to-noise ration for Y. Assuming that the 
detector noise is isotropic, Gaussian, uncorrelated among the detectors, and uncorrelated with the stochastic background signal, and greater than the stochastic background itself in all frequencies, it has been shown that \cite{Allen_Romano_90} the expectation value and variance of $Y$ become,
\begin{equation}
\mu_Y = <Y>= {T\over 2} \int_{-\infty}^\infty df \gamma(|f|) S_{GW} (|f|) {\tilde Q}(f) ,
\end{equation}
\begin{equation}
\sigma_Y^2 =<(Y-\mu_Y)^2 > \approx {T\over 4} \int_{-\infty}^\infty df S_1(|f|) |{\tilde Q}(f)|^2 S_2 (|f|),
\end{equation} 
where ${\tilde Q}$ is the Fourier transform of $Q$, $\gamma$ is the overlap reduction fucntion that chacterizes the reduction in sensitivity due to  time delay and relative orientation of two detectors and $S_1$ and $S_2$ are one-sided noise power spectra of the two detectors. The shape of the optimal filter function depends on the shape of the stochastic background via,
\begin{equation}
{\tilde Q} (f) \propto {\gamma(|f|) S_{GW} (|f|)\over S_1(|f|) S_2(|f|)}.
\end{equation}
For the case of constant $\Omega_{GW}(f) = \Omega_{GW,0}$, 
\begin{equation}
{\tilde Q} (f) \propto {\gamma(|f|) \over  |f|^3S_1(|f|) S_2(|f|)}.
\end {equation}
and signal-to-noise ratio  for $\mu_Y$ becomes
\begin{equation}
<\rho_Y> ={\mu_Y\over\sigma_Y} \approx {3H_0^2 \over 10 \pi^2} \Omega_{GW,0} \sqrt{T} \left[\int_{-\infty}^\infty df {\gamma^2(|f|)\over f^6 S_1(|f|) S_2(f|)}\right]^{1/2}.
\end{equation}
Unlike the single detector case, we can see that the signal-to-noise ratio increases with observing time. So far no detection of the stochastic background has been made yet, but the data taken during the first observing run of the advanced LIGO have been analyzed and  $\Omega_{GW,0} < 1.7\times 10^{-7}$ with 95 \% confidence \cite{Abbott_stochastic_17} was obtained. As the sensitivity of the detector and observing time increase the upper limit will go down significantly. The expected value of the  $\Omega_{GW,0} $ at around a few 100 Hz due to unresolved sources range from $10^{-10} \sim 10^{-8}$, and therefore the detection of the stochastic background could be made in the near future \cite{Abbott_stochastic_17}.

\section{Detected Sources}
As we have mentioned earlier in the introduction, so far seven gravitational wave events have been found mostly from LIGO: six black hole binaries and one binary neutron star. The distinction between black holes and neutrons stars is mostly based on the estimated masses of individual components, although the waveform of strong sources such as GW150914 seem to indicate that last cycle of the gravitational waves must have come when two objects became sufficiently close so that other types of stars cannot explain the observed waveforms. 

\begin{figure}
\includegraphics[width=10.0cm]{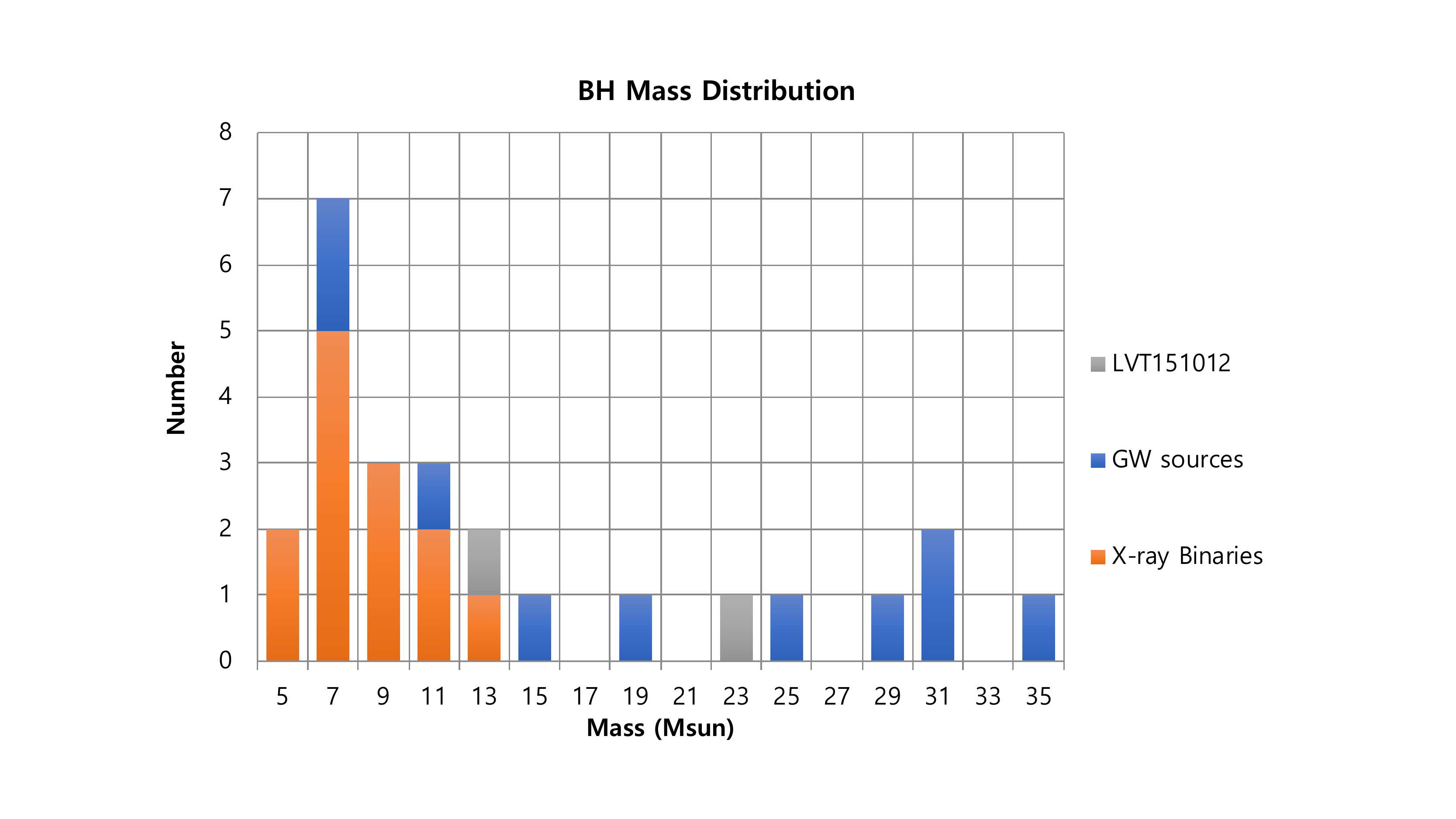}
\caption{The histogram of the masses of the black holes in X-ray binaries (red bars) and gravitational waves (blue bars). The X-ray binary black hole mass was taken from Wikipedia (${\tt https://en.wikipedia.org/wiki/Stellar\_black\_hole}$). Note that the black holes detected by the gravitational waves have wider range of masses with systematically higher mass than those in X-ray binaries.}

\end{figure}

\subsection{Black Hole Binaries}
The masses detected by the gravitational waves are somewhat different from those in X-ray binaries (see Fig. 4) in the sense that the range of masses is larger and average mass is higher. Note that the X-ray binaries can be observed in small distances and therefore those in Fig. 4 lie mostly in the Galaxy. On the other hand, the black hole binaries from gravitational waves are all from external galaxies. There is certainly a selection bias too since the black hole binaries with higher masses emit stronger gravitational waves and therefore those with higher masses can be preferentially detected. In the absence of prior knowledge on the distribution of black hole masses, however, such a selection effect is difficult to quantify.

The masses of the black holes formed through the stellar evolution are known to depend on the abundance of heavy elements since the stars lose mass significantly during the last phase of the evolution and the mass loss rate is higher for stars with higher abundance of heavy elements \cite{belczynski}. The exact dependence of the black hole mass on the chemical abundance is not well known. The fact that the black hole masses in X-ray binaries do not exceed $\sim 15 {\rm M_\odot}$ can be understood by the chemical elements of the progenitor stars since the Galactic X-ray binaries are relatively young populations.

The relatively higher masses of the black holes discovered by the gravitational waves are likely to have been formed in low metallicity environment. The chemical enrichment has taken place with cosmic time, and the black holes have been formed very early in the history of the universe. Another possibility is that the black holes are formed in low metallicity galaxies. Such galaxies are generally small in mass and it is difficult that majority of the black hole binaries from gravitational waves are formed in small galaxies.

If the black hole binaries are formed very early on, the orbital evolution must have taken very long time (nearly Hubble time). As we can see from equation (27). the merging time is a very sensitive function of semi-major axis, and therefore wide binaries take long time to merge (and even longer than Hubble time). For example, for the case of contact binaries, the separation is about $2 R_*$ where $R_*$ is the stellar radius. If the progenitor stars are  40 ${\rm M_\odot}$ stars, $R_*\approx 18 $R$_\odot$\cite{handbook}  and black hole binaries of 40 M$_\odot$ each with circular orbit of orbital radius of 36 R$_\odot$, the merging time is about $10^{12} yr$ which is about 70 times longer than Hubble time. Obviously, the binary evolution is much more complex. One star becomes a black hole first and the second star undergoes giant and superginat phase. The black hole coould well lie inside the companion star and the dynamical friction on the black hole by the stellar envelope could bring the black hole closer to the core of the companion. The separation between the black holes after the common envelope evolution is very uncertain.  

Another possibility is that the black holes are formed as a single ones in very early universe but became members of binaries by dynamical processes later.  There are two possible mechanisms for the formation of dynamical binaries: gravitational radiation capture and three-body processes. The gravitational radiation capture takes place when two black hole approach each other within very small distances so that more energy than kinetic energy at large distance is radiated away by gravitational waves. The cross section for gravitational radiation capture can be computed under parabolic approximation where the orbit is assumed to be parabolic and the amount of energy carried away by the gravitational waves is computed from 2.5PN approximation,

\begin{equation}
\sigma_{cap} = \pi r_{p,max}^2 \left[ 1 + {2G(m_1+m_2)\over r_{p,max} v_\infty^2}\right] \approx 17 {G^2 m_1 m_2 \eta^{-5/7}\over c^{10/7} v_\infty^{18/7}},
\end{equation}
where $\eta=m_1m_2/(m_1+m_2)^2$ is the symmetric mass ratio and $r_{p,max}$ is the maximum pericenter distance for the capture of encounters with relative velocity of $v_\infty$ between two point masses of $m_1$ and $m_2$, as given by \cite{quinlan_shapiro_89}
\begin{equation}
r_{p,max} =\left[{85\pi\sqrt{2} \over 12} {G^{7/2}\over c^5} {m_1m_2(m_1+m_2)^{3/2}\over v_\infty^2}\right]^{2/7}.
\end{equation}
The hyperbolic approximation is valid only for the case of $v_\infty\ll c$. Since the gravitational radiation capture is expected to take place in galactic nuclei star clusters where velocity dispersion is of order of a few hundred km/s, the parabolic approximation with 2.5PN should be an excellent approximation. Numerical simulations with full general relativity showed that the parabolic approximation leads to errors less than a few \% in capture cross sections even for $v_\infty$ of a few thousand km/s \cite{bae_etal_17}.
Such a process becomes efficient only in extremely dense stellar environments \cite{lee_95}\cite{hong_lee_15}.  The eccentricities of the captured binaries are extremely large and the time to merge is very small compared with other time scales relevant for the dynamics of the stellar systems. Some fraction of binaries are on substantially eccentric orbit when they enter the lower part of the  LIGO frequency band of about 30 Hz. The rate of binary formation is quite uncertain \cite{oleary_etal_09}\cite{hong_lee_15}, but likely to be less than 1 per ${\rm Gpc^{-3} year^{-1}}$. However, if we find any gravitational waves from black hole binaries on eccentric orbit, they are likely to be a result of gravitational radiation capture.

The binaries can be formed when three stars get close each other and one star get ejected by carrying large amount of energy so that the remaining star becomes bound. By using dimensional argument, one can show that the formation rate per unit volume for equal mass stars with $m$, stellar number density $n$ and the velocity dispersion $\sigma$ can be written
\begin{equation}
{dn_{3B}\over dt} = C{G^5m^5n^3\over \sigma^{9}}.
\end{equation}
 The strong dependence of formation rate on $\sigma$ makes this process very efficient in low velocity dispersion environments. Among dense stellar systems, globular clusters satisfy such a condition.
 
The dynamical formation of black hole binaries has been discussed in various context for a long time \cite{lee_95}. These are collectively called `dynamical binaries' while those formed via evolution of binary system are called `evolution binaries'. More recently, \cite{bae_etal_17} \cite{park_etal_17}  carried out detailed numerical simulations on the dynamical binaries in globular clusters. Following \cite{lee_95} they showed that the small number of black holes become a dominant component in short time after the formation in the central parts through dynamical friction since the black holes are the most massive components in the cluster after the massive stars are all evolved off. The black holes in the central parts of globular clusters interact closely each other and quickly form binaries via three-body processes. The binaries are rather wide so that the gravitational wave is not effective in bringing the binary components close. 

Instead, the interactions of binaries with single black holes bring the binaries closer since the triple interactions typically lead to the ejection of one member at higher speed. Such a process is called hardening. The hardening continues until the binaries are ejected from the clusters. The ejection takes place when the binary becomes sufficiently tight since typical recoil energy of the binaries and singles is proportional to the absolute value of the binding energy of the binary. 

Most of the binaries get ejected from the clusters before merger since the typical time scale for close encounters with single black holes is much shorter than the merging time. However, as the black holes become depleted, the interaction would be rare and those binaries that are formed in the later phase will merge inside the cluster, as shown in Fig. 5. It is evident that a small fraction of black hole binaries could merge inside clusters.

\begin{figure}
\includegraphics[width=10.0cm]{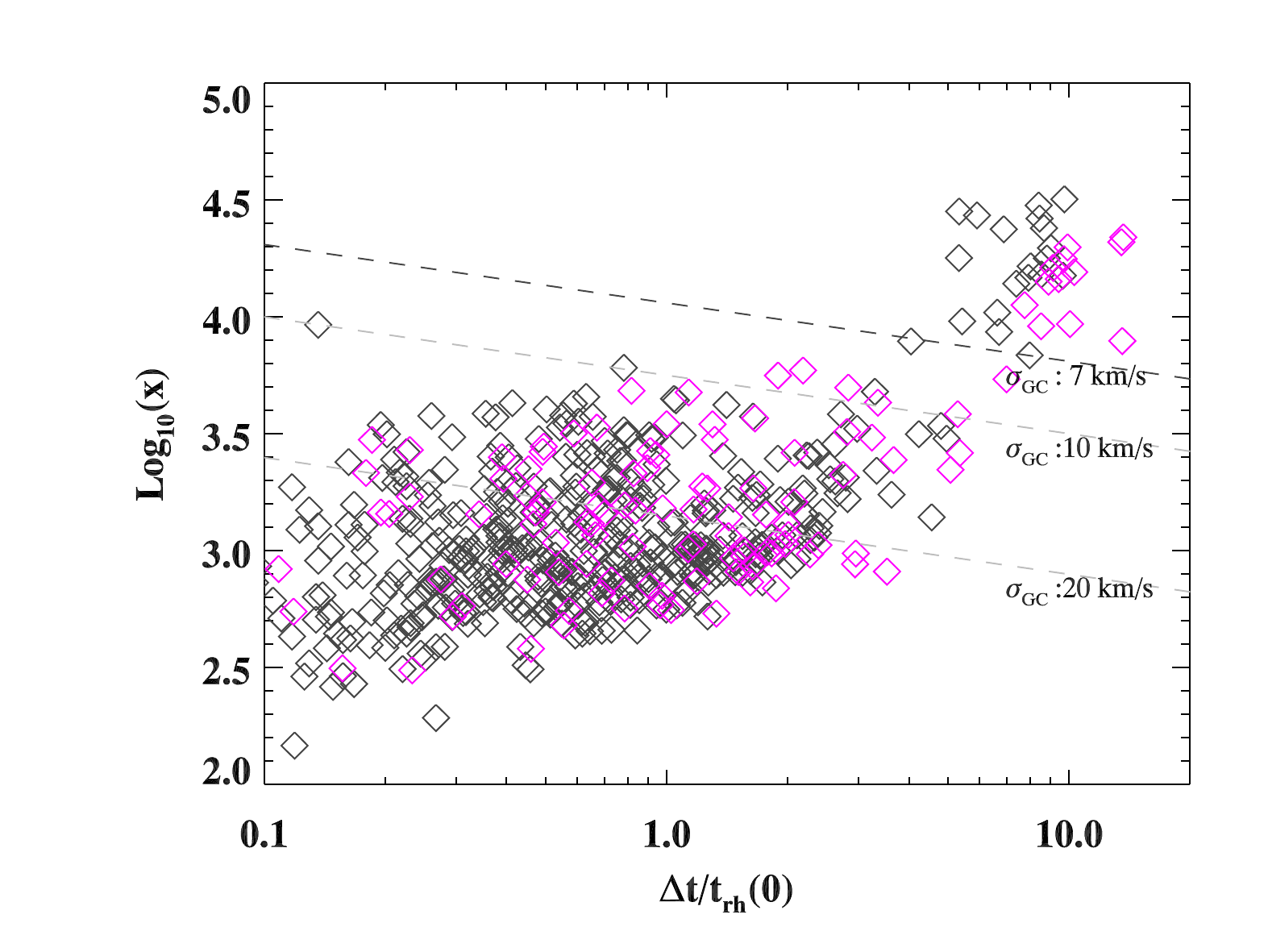}
\caption{The time gap between the formation and ejection of binaries versus the hardness of the binaries based on N-body simulations. The dashed lines indicate the merging time for clusters with different velocity dispersion. Any binaries lying above these lines are likely to merge before they get ejected. The fraction of such binaries is expected to be less than 10\%.
}
\end{figure}
The numerical calculations predicted that the dynamically formed binaries in globular clusters could outnumber neutron star binary mergers in the range of advanced LIGO. Obviously the black hole binaries form

\subsection{Merging Event of Neutron Star Binary}
On August 17, 2017, a new type of the gravitational wave event was discovered by the LIGO. Unlike previous events, a gravitational wave lasted almost for 1 minutes until the last cycle of the gravitational wave at frequency of $\sim$ 800 Hz\cite{GW170817}. The total mass was estimated to be much smaller than those of the previous events and the information regarding the sky location of the event was spread to the electromagnetic partners in the world. The Fermi Gamma Ray Burst Monitor (GBM)\cite{GBM} and {\it |INTEGRAL}\cite{Integral}  also detected the burst of gamma-ray about 1.7 seconds after the end of the gravitational wave signal. The gamma-ray event was classified as short gamma-ray burst. The accuracy of the sky location was much improved when the data from the Virgo were analyzed together, even though the Virgo was not able to detect the gravitational wave signal. Virgo joined the observing run on August 1 with relatively lower sensitivity than LIGO. The fact that the Virgo could not see the signal meant that large portion of the sky location with two LIGO detector only could be eliminated. The 90\% confident sky location was initially about  $\sim$ 31 square degrees (later improved to 28 square degrees by more  careful analysis) and the electromagnetic partners used such a constraint to search for the electromagnetic counterparts. About 10.9 hours later, optical counterpart was discovered by Swope supernova survey \cite{swope} in an elliptical galaxy NGC4993. Many ground-based and space-based telescopes followed up the observation of this source\cite{}. The Korean team of Seoul National University and Korea Astronomy and Space Science (KASI) also observed the photometric variation of the optical counterpart using several telescopes including KMTNet (Korea Microlensing Telescope Network)\cite{KMT} which is composed of three identical telescopes located at Las Campanas in Chile, Siding Spring Observatory in Australia, and South African Astronomical Observatory in South Africa. The optical emission is interpreted as due to heating by radio active decay of r-process elements as predicted by kilonova \cite{kilonova_1} \cite{kilonova-2} models.

The electromagnetic followup covered almost all possible wavelengths, from radio to X-ray. Our understanding of the short gamma-ray burst has been advanced from the nearly simultaneous detection of gravitational waves and gamm-ray emission. The short gamma-ray burst has been suspected to originate from the merging of neutron star binary but no observational confirmation was made until GW170817/GRB170817A event. The X-ray emission was observed about days after the gamma-ray burst \cite{korea_em_followup} by Chandra. Detection of radio emission at 3 and 6 GHz has been made with the Very Large Array (VLA) 16 days after the merger \cite{radio}. The X-ray and  radio emissions are thought to be due to the interactions of the jet with surrounding medium.  By combining these observations, it was concluded that merger of two neutron stars generated a hot explosion, causing the rapid production of r-process elements in the merger ejecta. The host galaxy is located at around 40 Mpc, one of the closest host for gamma-ray bursts, but the gamma-ray was found to relatively weak. This means that we are viewing the gamma-ray burst off axis.  The infrared emission is dominated by the neutron rich ejecta while the optical emission is mostly due to neutron free wind along the polar axis.

Even though the connection between the neutron star merger and the short gamma-ray burst is now well established through the observation of GW170817/GRB170817A many questions remain.. It is not clear whether there was a jet since the gamma-ray emission was significantly weaker than any of the previous sGRBs. This was interpreted as the off-axis jet, but recent observation at X-ray showed the significant brightening 109 days after the merger. Such a brightening is not possible with simple jet model\cite{brightening}, suggesting the failure of the jet to break out the ejecta cloud. Instead the cocoon model is more preferred to explain the brightening of both radio and  X-ray emission. 

The host galaxy NGC 4993 has been also studied in detail\cite{im_etal_17}. Based on the deep image obtained by stacking large number of images taken during the followup observation campaign at many different wavelengths, morphological, structural and photometric properties have been analyzed.  The surface brightness profile of NGC 4993 found to  follow those typical elliptical galaxies. There is The spectral energy distribution of NGC 4993 was used to probe the star formation history. By fitting the observed spectral energy distribution covering from 1.5 nm to 20 $\mu$m to the population synthesis models, mean stellar age is estimated to be greater than $\sim$ 3 Gyrs. This means that the neutron star binary was formed more than 3 Gyrs ago, indicating that the neutron star binary may have formed no later than 3 Gyrs ago. The deep Hubble Space Telescope image in 606 nm for the inner parts of the galaxy shows trace of dust lane, implying the possible minor merger with a gas rich galaxy recently, although there is no evidence for recent star formation activity. 

The luminosity distance to the gravitational wave source can be estimated and the Hubble constant $H_0$ can be measured if the redshift of the host galaxy is known. Such a possibility has been first pointed out by Schutz \cite{standard_siren}. The luminosity distance estimation inevitably contains uncertainity mostly due to the uncertainty in the inclination angle of the orbital plane. The estimated Hubble parameter using the gravitational wave data  for GW180717 was $H_0 = 70.0_{-8.0}^{+12.0}$ km/s/Mpc (68.3\% credible range \cite{hubble}). Obviously current estimation is compatible both with tother measurements such as supernova \cite{hubble_sn} and the cosmic microwave background\cite{hubble_planck}. However, with more detections of binary neutron star mergers and their hosts become the accuracy of the $H_0$ will improve.

\section{Future Detectors and Prospects}

The gravitational wave is now well established as a tool to study astrophysical objects.  The advanced LIGO and Virgo will add more sources in the forthcoming observing runs with better sensitivity. With accumulation of more data, we will be able to have better understanding on the population of black holes and neutron stars. Other types of sources than compact binary mergers will be detetced as the detector sensitivity increases. 

More detector will become available in the near future. KAGRA, a Japanese detector located in the underground of Kamioka  mountain will join ground based detector network soon \cite{kagra} and LIGO India \cite{ligo-india}is now under construction. If these detectors operate with LIGO and Virgo at their design sensitivities, a significant fraction of gravitational wave signals will be localized to a few square degrees by gravitational wave observations alone \cite{LVK}. This will increase the possibility of identifying host galaxies significantly. 

The gravitational waves at significantly lower frequencies than the LIGO/Virgo band can be covered by the space-based detectors in the future. Laser Interforometer Space Antenna (LISA\url{https://www.elisascience.org} ) is a space observatory in space  to explore the gravitational wave universe. The freuency range to be covered by LISA is  between 0.1 mHz to 1 Hz. The sources for the detectors like LISA include white dwarf binaries, stellar mass black hole binaries in earlier stage of the inspiral, binaries composed of intermediate mass black holes and inspiral of stellar mass black black holes onto supermassive black holes (also known as extreme mass ration inspiral or EMRI).

Several concepts of detectors for mid-frequencies (0.1$\sim$ 10 Hz)\cite{harms_etal_12} where neither  ground-based detectors nor LISA is sensitive. In these frequencies, the seismic and Newtonian gravity gradient noses become serious. In particular Newtonian noise cannot be mitigated easily since the Newtonian forces cannot be blocked. The three-dimensional detector proposed in \cite{Paik_etal_16} may be able to cancel Newtonian noises using full tensor components of gravity gradient in 3-dimension \cite{harms_paik}. The scientific targets in this frequency range include inspiral of stellar mass black hole onto  an intermediate mass black hole (also called intermediate mass ratio inspiral or IMRI) and binaries composed of intermediate mass black holes. The stellar mass black hole binaries can also be detected about a week before the merger can be seen by detectors in the mid-frequencies \cite{paik_etal_17}. Atom interferometry as a nw detection technique for gravitational waves in mid-frequencies was proposed  \cite{Dimopoulos}. In order to avoid noises from seismic activities, very sensitive space antenna named as DECI-hertz Interferometer Gfavitational Wave Observatory (DECIGO) has been proposed also to fill the gap between the ground-based detectors and the LISA \cite{sato_etal}.  

The future detectors will enrich our understanding of the universe significantly by adding new types of sources.For example, intermediate mass black holes are only suspected from evidneces such as the motion of stars in the central parts of globular cluters or from X-ray luminosities of certain sources. The ultimate proof should come from the derection of gravitational waves. Furthermore, there might be many astrophysical objects waiting to be discovered by the  observation of gravitational waves. 

\begin{acknowledgments}
The author acknowledges Hyun Kyu Lee, Gungwon Kang and Sukanta Bose for the careful reading of the manuscript.
\end{acknowledgments}

\end{document}